\documentclass[conference]{IEEEtran}
\IEEEoverridecommandlockouts
\usepackage{cite}
\usepackage{subfig}
\usepackage{amsmath,amssymb,amsfonts}
\usepackage{algorithmic}
\usepackage{graphicx}
\usepackage{textcomp}
\usepackage{xcolor}
\usepackage{booktabs}
\usepackage{siunitx}
\usepackage{doi}

\def\BibTeX{{\rm B\kern-.05em{\sc i\kern-.025em b}\kern-.08em
    T\kern-.1667em\lower.7ex\hbox{E}\kern-.125emX}}
\begin{document}

\title{Concept of a System-on-Chip Research Platform Benchmarking Interaction of Memristor-based Bio-inspired Computing Paradigms
\thanks{This research work within the NEUROTEC II project is funded by the German Federal Ministry of Education and Research (BMBF) with grant number 16ME0398K.}
}
\makeatletter
\newcommand{\newlineauthors}{%
  \end{@IEEEauthorhalign}\hfill\mbox{}\par
  \mbox{}\hfill\begin{@IEEEauthorhalign}
}
\makeatother
\author{
    \IEEEauthorblockN{
        Christian Grewing\IEEEauthorrefmark{1},
        Arun Ashok\IEEEauthorrefmark{1},
        Sabitha Kusuma\IEEEauthorrefmark{1},
        Michael Schiek\IEEEauthorrefmark{2},
        André Zambanini\IEEEauthorrefmark{1}, and
        Stefan van Waasen\IEEEauthorrefmark{1}\IEEEauthorrefmark{3}
    }\\
    \IEEEauthorblockA{
        \IEEEauthorrefmark{1}Peter Grünberg Institute -- Integrated Computing Architectures (ICA $|$ PGI-4),\\
        Forschungszentrum Jülich GmbH,
        52425 Jülich, Germany\\
        \IEEEauthorrefmark{2}Peter Grünberg Institute --
        Neuromorphic Compute Notes (PGI-14),
        Forschungszentrum Jülich GmbH,
        52425 Jülich, Germany\\
        \IEEEauthorrefmark{3}Faculty of Engineering, 
        Communication Systems, 
        University of Duisburg-Essen, 
        47057 Duisburg, Germany\\[1mm]
        Email: c.grewing@fz-juelich.de
    }
}

\maketitle
\begin{abstract}
  A system architecture is suggested for a System on Chip that will combine several different memristor-based, bio-inspired computation arrays with inter- and intra-chip communication. It will serve as a benchmark system for future developments. The architecture takes the special requirements into account which are caused by the memristor co-integration on commercial CMOS structures in a post processing step of the chip. The interface considers the necessary data bandwidth to monitor the internal Network on Chip at speed and provides enough flexibility to give different measurement options.
\end{abstract}


\section{Introduction}
The ever-increasing demand for computational performance rises a particular issue with electrical power consumption. One significant bottleneck in certain operations is caused by the van Neumann architecture and is especially relevant in artificial intelligence applications \cite{Ricci17thermo}. Bio-inspired computation paradigms like Computing in Memory (CiM), Content addressable Memory (CaM), Spiking Neural Networks (SNN), or Probabilistic Computing (PC) are promising concepts to overcome this bottleneck in standard computing architectures \cite{aguirre_hardware_2024}. All of these have been studied in detailed, based on numerical models and hardware implementations, many utilizing the capabilities of novel memristive devices \cite{Bao22CiM,Li20CaM}. However, these approaches only address particular computational steps and a widespread usage requires a versatile System on Chip (SoC) to allow the interaction of these approaches. We propose a system architecture including all of these approaches on one chip to allow a quantitative prediction of performance, energy consumption, and footprint in contrast to standard, pure-CMOS implementations. The components are connected through a Network on Chip (NoC) and a chip bridge for die-to-die connections, enabling a flexible, scalable, and efficient interaction between these. Our system concept successfully overcomes today's challenge that most memristive devices are not included in the commercial CMOS manufacturing flow and need to be added to the chip structures in a post processing step. The paper describes the considerations and constraints leading to the presented system as part of the research for a demonstrator system.

\begin{figure}[ht]
  \centering
  \includegraphics[width=\linewidth]{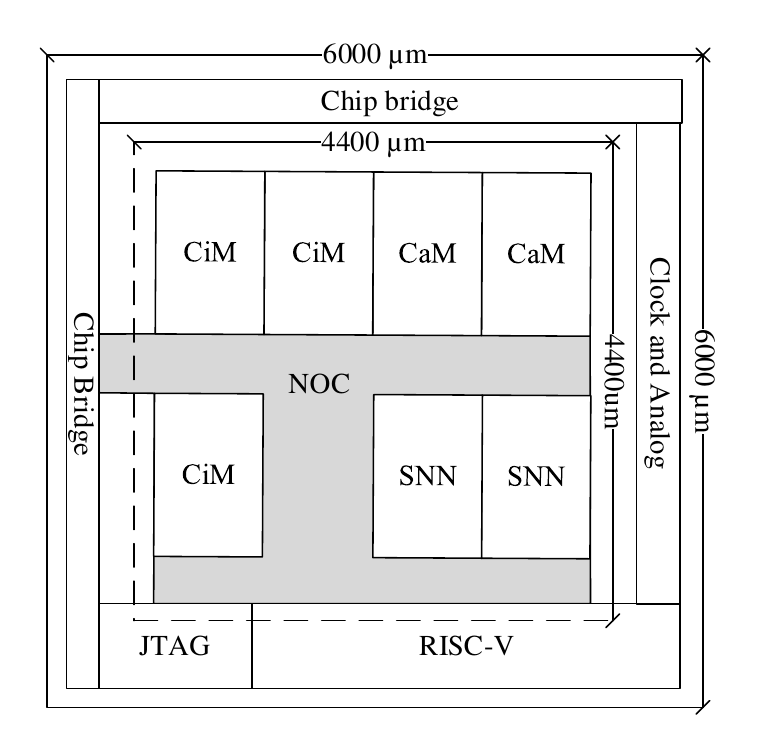}
    \caption{Architecture and floorplan of the proposed SoC.}
\end{figure}

\section{Architecture and Floorplan}
The chip contains seven Computing Arrays (CAs) as the bio-inspired computing core, joined with auxiliary circuits to support with the versatility of the system. The CAs are all built with a memristor-based crossbar that is interfaced through mixed-signal components and a local controller and local memory to allow independent computation operations. Besides the novel computation cores, a classical 32-bit RISC\nobreakdash-V processor with on-chip SRAM is foreseen to control and provide data to the seven CAs. All eight elements are interconnected through a 32-bit NoC and configured through a JTAG programming interface. External access to the system is provided through a chip bridge, allowing at-speed readout of the traffic in the NoC and demonstrating the system's scalability by connecting two chips. Additionally, the RISC-V can be directly accessed through a dedicated AXI interface. 


\section{Requirements}
\subsection{Technology and Memristor Fabrication}
A major goal of the demonstrator is to show the possibility of integrating memristors into a scalable system, where the CMOS chip will be fabricated in a commercial technology and the memristors will be fabricated on top in-house. 
The chosen technology is a 28nm bulk CMOS process as it enables a high signal bandwidth and its feature size is small enough for a highly integrated and scalable SoC. Further, in our experience the design and layout effort for the mixed-signal circuitry is significantly lesser compared to lower technology nodes. 
Since the connection pads for the memristor arrays need to be small with a thin top metal layer requiring a post-processing to remove the thick aluminium structures.
The post-processing requires a minimum die size of \qtyproduct{6 x 6}{\milli\meter} to enable post-processing with etching, lithography, and depositing steps which show undesired edge effects if around \SI{0.8}{\milli\meter} close to the chip boundary. Hence, all memristor circuits should be in the center square of \qtyproduct{4.4 x 4.4}{\milli\meter}. 
While naked dies will allow direct characterization of control structures for the memristor processing yield, most dies will be packaged to simplify handling. The packaging as well as the high density of memristors makes the individual memristor only accessible through CMOS circuits. However, memristors require an initialization step called electroforming with around \SI{3}{\volt} which demands the technology option for \SI{3.3}{\volt} as an input-output (IO) voltage in the padframe.


\subsection{Package and Bonding}
As a package, a Very thin Quad Flat No leads package (VQFN) with a ground connection at the bottom of the package is chosen. The chip fits well into larger variants, which means that the longest bonding wires should not exceed \SIrange{2.5}{3}{mm}, limiting the wire inductivity and making high data bandwidths possible. The substrate is glued on the island, providing a very good ground connection and low thermal resistance. The ground pads of the chip are connected with down-bonds to the package bottom. 


\begin{table}[htb]
\centering
\caption{Estimation of Current Consumption}
\begin{tabular}{llcc}
\toprule
    \textbf{Supply} &\textbf{\textit{Block}}& \textbf{\textit{Current}}&\textbf{\textit{max. Clock}} \\ 
\midrule
    \multicolumn{4}{l}{\hspace{2mm}\textit{Computing Array Circuits}}\\
    VDD\_CORE& SRAM (\SI{32}{\kilo\byte}) \textit{per CA}	&\SI{18.9}{\milli\ampere}& \SI{1}{\giga\hertz} \\
    VDD\_CORE &local controller \textit{per CA}&	\SI{10}{\milli\ampere}&	\SI{1}{\giga\hertz}\\
    VDD\_ANA &analog component \textit{per CA}&	 \SI{3.5}{\milli\ampere} 	&	\SI{100}{\mega\hertz}\\[1mm]
    \multicolumn{4}{l}{\hspace{2mm}\textit{Common Digital Circuits}}\\
    VDD\_CORE&RISC-V	&\SI{10}{\milli\ampere}&	\SI{500}{\mega\hertz}\\
    VDD\_CORE&	SRAM ($2\times\SI{64}{\kilo\byte}$)&\SI{51.6}{\milli\ampere}&		\SI{500}{\mega\hertz}\\
    VDD\_CORE&AXI Interface&	\SI{2}{\milli\ampere}	&	\SI{100}{\mega\hertz}\\
    VDD\_CORE&	Network on Chip &\SI{10}{\milli\ampere}	& \SI{1}{\giga\hertz}\\
    VDD\_CORE&	Chip Bridge&\SI{10}{\milli\ampere}	&\SI{1}{\giga\hertz}\\[1mm]
    \multicolumn{4}{l}{\hspace{2mm}\textit{IO Structures}}\\
    VDD\_LVDS&Output Chip bridge	&\SI{95}{\milli\ampere}&\SI{1}{\giga\hertz}\\
    VDD\_CORE&Input Chip bridge	&\SI{12}{\milli\ampere}&\SI{100}{\mega\hertz}\\
\bottomrule
\end{tabular}
\label{tab:current}
\end{table}

\subsection{Power Supply}
The chosen technology offers a \SI{0.9}{\volt} core voltage and a \SI{3.3}{\volt} IO voltage. For the general functionality of the memristor, the \SI{0.9}{\volt} is sufficient \cite{bende2023experimental} for the computing mode. It is also enough for the mixed-signal circuitry such as Analog-to-Digital Converters (ADC), Digital-to-Analog Converters (DAC), analog multiplexer, biasing circuits, and clock buffers. A separate supply pad for each crossbar is beneficial, so that the individual current consumption can be measured constantly and the parasitic coupling between CAs is reduced. The \SI{3.3}{\volt} level is used in the padframe for the ESD protection circuitry as well as for memristor forming and manipulation, i.e. SET and RESET operations to change the conductance values. 
The main factor for the current consumption of the digital supplies is the frequency and complexity of the circuit, so the whole design is segmented into smaller parts. For optimization, current estimations are assigned to each block based on previous designs and accumulated experience. The most concrete estimation can be made for the SRAM as this scales directly with the size and frequency of the memory. Even though most use cases won't require a full utilization and some elements will be shut off, a worst-case estimate is required. Table \ref{tab:current} lists the collection of all estimates.
Finally, the main limitation needs to be identified to estimate the number of required pads. Here, the suppliable current is mainly restricted by the pad and ESD structures embedded in the CMOS circuit, limiting the current to a low, to-digit Milliampere rating per pad. This leads to the given number of pads in Table \ref{tab:supply}.

\begin{table}[htb]
\centering
\caption{Interface Definition of Supply Signals}
\begin{tabular}{llc}
\toprule
    \textbf{Name} & \textbf{\textit{Purpose}}& \textbf{\textit{Pads}} \\
\midrule
    \multicolumn{3}{l}{\hspace{2mm}\textit{Digital Supplies}}\\
    VDD\_CORE& \SI{0.9}{\volt} supply of digital part& 20  \\
    VSS\_CORE& Low supply of digital part& 20 \\[1mm]
    VDD\_CLK& \SI{0.9}{\volt} supply of clock buffer & 1  \\
    VSS\_CLK& Low supply of clockbuffer& 1 \\[1mm]
    VDD\_IO& \SI{3.3}{\volt} supply of digital section of padframe& 2  \\
    VSS\_IO& Low supply of digital part of the padframe& 2 \\[1mm]
    VDD\_LVDS& \SI{1.4}{\volt} supply of LVDS& 13  \\
    VSS\_LVDS& Low supply of LVDS& 13  \\[2mm]
    \multicolumn{3}{l}{\hspace{2mm}\textit{Analog Supplies}}\\
    VDD\_ANA& \SI{0.9}{\volt} supply of analog part of the CA& 7  \\
    VSS\_ANA& Low supply of analog part of the CA& 7 \\
    VDDA\_ESD& \SI{3.3}{\volt} supply of analog section of padframe& 2  \\
    VSSA\_ESD& Low supply of analog part of the padframe& 2 \\
    VDDA\_EF& \SI{3.3}{\volt} for electroforming, set and reset of  & 7  \\
\bottomrule
\end{tabular}
\label{tab:supply}
\end{table}

\begin{table}[htb]
\centering
\caption{Interface Definition}
\begin{tabular}{llcc}
\toprule
    \textbf{Name} & \textbf{\textit{Purpose}}& \textbf{\textit{Speed}}& \textbf{\textit{Pads}} \\
    && \textit{per line}&\\
\midrule
    \multicolumn{4}{l}{\hspace{2mm}\textit{Clock}}\\
    CLK 3:0& Clock inputs &\SI{1}{\giga\hertz}& 4  \\[2mm]
    \multicolumn{4}{l}{\hspace{2mm}\textit{LVDS Pairs}}\\
    CBTXDAT 15:0& Chip bridge TX data \textit{(TX)} &\SI{2}{\giga\bit/\second}& 32  \\
    CBTXADD 1:0& Chip bridge address \textit{(TX)} &\SI{2}{\giga\bit/\second}& 4 \\
    CBTXVAL& Chip bridge data valid \textit{(TX)} &\SI{2}{\giga\bit/\second}& 2 \\
    CBTXREA& Chip bridge TX ready \textit{(RX)} &\SI{2}{\giga\bit/\second}& 2 \\[2mm]
    \multicolumn{4}{l}{\hspace{2mm}\textit{Single-Ended TTL Lines}}\\
    CBRXDAT 7:0& Chip bridge RX data &\SI{100}{\mega\bit/\second}& 8 \\
    CBRXADD& Chip bridge RX address &\SI{100}{\mega\bit/\second}& 1 \\
    CBRXVAL& Chip bridge RX data valid &\SI{100}{\mega\bit/\second}& 1 \\
    CBRXREA& Chip bridge RX data ready &\SI{100}{\mega\bit/\second}& 1 \\
    AXI 14:0 & Processor control &\SI{100}{\mega\bit/\second}& 15 \\
    JTAG 4:0& Control of register settings &\SI{10}{\mega\bit/\second}& 5 \\
    INTRPT 1:0 & Asynchronous interrupt&& 2 \\[2mm]
    \multicolumn{4}{l}{\hspace{2mm}\textit{Test Signals}}\\
    SCAN\_EN & Enable scan chain&& 1 \\
    TESTAC& Analog test signal& \SI{500}{\mega\hertz}& 7 \\
    TESTDC& Analog test signals& DC& 7  \\
\bottomrule
\end{tabular}
\label{tab:IO}
\end{table}

\subsection{Interfaces of the Chip}
The main interaction with the chip will be through digital interfaces. The most important will be the chip bridge, which allows at-speed output from the NoC through 22 Low Voltage Differential Signaling (LVDS) pairs and grants direct access to the CAs through additional 11 single-ended Transistor-Transistor Logic (TTL) lines. The general programming and slow control interface of the chip is provided through JTAG while the RISC-V processor may be accessed directly through a dedicated, 15-bit wide AXI stream interface. The latter serves as a required interface to an external processor, also part of the demonstrator system. Finally, two asynchronous signals act as digital interrupts. One is an output to be used as a prompt, for instance, to signal a ready state or to show soft errors like memory overflows. The other is an input signal to select one of two configuration register banks or other immediate control features. This enables a fast, asynchronous switching between two different states, adding flexibility in the use of a chip. The clock of the chip can be fed into four different pins in order to enable slight clock skewing if necessary. The interface is also summarized in Table \ref{tab:IO}


\subsection{Testability}
The testability of the system can be classified into production, characterization and system verification tests. The first will be applied before the post-processing steps to avoid performing intricate steps on faulty devices. Due to the extensive digital circuitry on the chip, CMOS manufacturing faults are becoming more likely and a comprehensive scan chain test is required. At a minimum, a dedicated enable signal is needed while other in- and outputs can be multiplexed with other pads. Since this test will be performed before the packaging, the access is restricted to direct contact with the chip pads.


The chip bridge is used to monitor the data on the NoC with enough data bandwidth for typical applications for system verification. The pinning needs also to provide enough flexibility to make different measurement points accessible or enable variations in the setup for characterization of individual parts.  Additionally, dedicated analog test pads for alternating and stable signals are included to access internal measurement points from various sources. Further, automated test routines like memory built-in self-tests will be included to reduce sources of undetected malfunction and improve qualification speed.



\section{Summary and Outlook}
 A system architecture of a SoC including all interfaces was presented to demonstrate the heterogeneous integration of different bio-inspired computation paradigms. The supply of the digital part and the LVDS would require 36 pins and corresponding 36 pads for the ground downbonds. 7 analog supplies with one for each CA and 9 pads at the 3.3V for ESD, electroforming, SET and RESET of the memristor are needed. The analog test connections require 14 pads. The NoC can be monitored via the chip bridge with 20 LVDS Transmitters and 1 LVDS Receiver. The input to the chip bridge comes with 10 TTL inputs and 1 TTL output. Further control interfaces are the 15-bit AXI interface, the 5 pads JTAG, and 2 interrupt signals, completed with 4 clock inputs to allow skewing if necessary. An enable signal for the boundary scan is added as well. In total, this requires 46 supply pins, 46 ground pads with down bonds, 14 analog test pins, 42 pins for the LVDS connection, and 31 TTL digital IO pins. The package needs 140 pins with external access and 46 down-bonds for a shared ground connection in a VQFN.

The corresponding chip is currently being developed while partial features are included in smaller, pure-CMOS demonstrators to allow the characterization of crucial circuits. The first measurement results of the full demonstrator setup are expected in the summer of 2025.

\section{Acknowledgement}
The SoC is developed for the demonstrator project of NEUROTEC II (16ME0398K), carried out at Forschungszentrum Jülich and funded by German Federal Ministry of Education and Researc (BMBF).

\bibliographystyle{ieeetr}
\bibliography{mybib}{}

\end{document}